\documentclass[a4paper]{article}

\usepackage{amsmath}
\usepackage{amsthm}
\usepackage{amssymb}

\newcommand{\Bra}[1]{\left<#1\right|}
\newcommand{\Ket}[1]{\left|#1\right>}

\newcommand{\Set}[1]{\{#1\}}

\newcommand{\Norm}[1]{\left\|#1\right\|}

\DeclareMathOperator{\comp}{comp}
\DeclareMathOperator{\compeps}{\comp_\epsilon}

\DeclareMathOperator{\compqeps}{comp^{q}_\epsilon}

\DeclareMathOperator{\compqpeps}{\comp^{\text{q,phase}}_\epsilon}
\DeclareMathOperator{\compqbeps}{\comp^{\text{q,bit}}_\epsilon}

\DeclareMathOperator{\errqn}{e^{q}_n}

\DeclareMathOperator{\id}{id}

\DeclareMathOperator{\spanSpace}{span}

\newtheorem{defi}{Definition}
\newtheorem{theo}[defi]{Theorem}
\newtheorem{lemm}[defi]{Lemma}
\newtheorem{corr}[defi]{Corollary}

\newtheorem{prop}[defi]{Proposition}

\begin{document}

\title{The Power of Various Real-Valued Quantum Queries}

\author{Arvid J. Bessen\footnote{bessen@cs.columbia.edu}}

\date{\today}

\maketitle
\begin{centering}
{Universit\"at Kaiserslautern\\
Fachbereich Informatik\\
Postfach 3049\\
D-67663 Kaiserslautern}

\end{centering}

\begin{abstract}
    The computation of combinatorial and numerical problems on quantum computers is often much faster than on a classical computer in numbers of queries.
  A query is a procedure by which the quantum computer gains information about the specific problem.

  Different query definitions were given and our aim is to review them and to show that these definitions are not equivalent.
  To achieve this result we will study the simulation and approximation of one query type by another.
  While approximation is ``easy'' in one direction, we will show that it is ``hard'' in the other direction by a lower bound for the numbers of queries needed in the simulation.
  The main tool in this lower bound proof is a relationship between quantum algorithms and trigonometric polynomials that we will establish.

\end{abstract}

\section{Introduction to Quantum Queries}

Since Grover's discovery of the quantum search algorithm \cite{gro-96a}, the notion of \emph{query complexity} has played an important role in quantum computation (for a thorough introduction to quantum computation see Nielsen, Chuang, \cite{nie-chu-00}).
Grover studied the problem to find an index $j \in \{ 0, 1, \ldots, N-1 \}$ so that for a function
\[ f : \{0, \ldots, N-1 \} \rightarrow \{ 0, 1 \} \]
$f(j) = 1$ holds.
He constructed a quantum algorithm that needs $\mathcal{O} ( \sqrt{N} )$ evaluations of $f$, if exactly one index $j$ with $f(j) = 1$ exists, while an algorithm on a classical computer would need $\Theta ( N )$ evaluations.
His quantum evaluation of $f$ was given by the transformation
\begin{eqnarray*}
  \Ket{j} \mapsto \begin{cases}
    \Ket{j} & \text{ if } f(j) = 0 \\
    - \Ket{j} & \text{ if } f(j) = 1
  \end{cases}
\end{eqnarray*}
This query can easily be seen to be equivalent to the query
\begin{eqnarray}
  \label{eqn:bool-query}
  \Ket{j} \Ket{b} \mapsto \Ket{j} \Ket{b \oplus f(j)}
  = \Ket{j} \Ket{b + f(j) \mod 2}
\end{eqnarray}
which was used by Beals et al. \cite{bea-buh-cle-mos-wol-98} and Nayak and Wu \cite{nay-wu-99}. For this query they proved important lower bounds for many problems that involve Boolean queries.

All these queries refer to the following model of computation.
Let $Q_f$ denote a query transformation for a fixed function $f$, let $U_k$ denote arbitrary unitary transformations not depending on $f$, and let $\Ket{\psi}$ be a starting state.
Consider the quantum algorithm
\begin{equation}\label{eqn:quantum-algorithm}
  U_n Q_f U_{n-1} \ldots U_1 Q_f U_0 \Ket{\psi}.
\end{equation}
It returns a resulting state $\Ket{\chi_f}$ after measurement with probability $p_{\chi_f}$.
We identify this state $\Ket{\chi_f}$ with its index $\chi_f$ and map it to the solution space by a computable mapping $\varphi$ not depending on $f$ on a classical computer.

We briefly recall some definitions from information-based complexity (see the introductions and surveys Traub, Werschulz \cite{tra-wer-98}, and Traub, Wasilkowski, Wo{\'z}niakowski \cite{tra-was-woz-88}).
These were transferred to quantum computers by Heinrich, \cite{hei-01}, whose approach we follow.
Let the correct solution to every problem $f$, $f$ from the problem class $F$, be given by a mapping $S : F \rightarrow G$.
If the algorithm returns a $\varphi ( \chi_f ) \in G$ for every input $f \in F$ with probability $p_{\chi_f}$, so that
\[ \sum_{ \chi_f : \| S(f) - \varphi (\chi_f) \| < \epsilon} p_{\chi_f} \geq \frac{3}{4}, \]
it is said to solve the problem for the problem class $F$.
The best precision $\epsilon$ that can be achieved by a quantum algorithm, which uses only $n$ queries, is called the \emph{$n$-th minimal quantum query error $\errqn(S,F)$} .
Conversely the \emph{quantum query complexity $\compqeps(S,F)$} is the minimal number of queries that have to be used in a quantum algorithm to get the desired result with precision $\epsilon$.

With this formal framework we are able to deal with numerical problems, if we define what the query $Q_f$ is in this case.
Since numerical queries have more than just two possible outcomes we have to extend our notion of a query.
We will follow the formal approach of Heinrich to discuss the different queries.
Suppose we want to query a function $f : D \rightarrow K$.
To ease our discussion let us restrict to $K \subseteq [0,1]$.
We have to choose a mapping $\tau : \{0, 1, \ldots, 2^n-1\} \rightarrow D$, decoding the input, and a mapping $\beta$ encoding the output.

There are two different approaches to encode a value in $[0,1]$ in a quantum computer.
The first approach was established by Abrams and Williams \cite{abr-wil-99}, and Novak \cite{nov-01}, who proposed the following query
\begin{equation}
  \label{eqn:phase-query}
  \begin{split}
    Q^{\text{phase}}_f \Ket{j} \Ket{0}
    & = \sqrt{1 - \beta(f(\tau(j)))} \Ket{j} \Ket{0} + \sqrt{\beta(f(\tau(j)))} \Ket{j} \Ket{1} \\
    Q^{\text{phase}}_f \Ket{j} \Ket{1}
    & = - \sqrt{\beta(f(\tau(j)))} \Ket{j} \Ket{0} + \sqrt{1 - \beta(f(\tau(j)))} \Ket{j} \Ket{1}
  \end{split}
\end{equation}
This query continuously changes between the two states $\Ket{0}$ and $\Ket{1}$.
We will refer to this query as the \emph{phase query} since it can be viewed as a (relative) phase rotation in the space spanned by $\Ket{0}$, $\Ket{1}$.
In \cite{abr-wil-99} the mapping $\beta$ was chosen as 
\begin{eqnarray*}
  \beta : [0,1] & \rightarrow & [0,1] \\
  x & \mapsto & x^2
\end{eqnarray*}
while in \cite{nov-01} $\beta = \id$.

Another approach to extend the Boolean query was chosen by Heinrich, \cite{hei-01}.
He encoded the values in $[0,1]$ via $\beta$ to $\{0,1,\ldots,2^m-1\}$ and defined:
\begin{eqnarray}
  Q^{\text{bit}}_f \Ket{j} \Ket{x}
  = \Ket{j} \Ket{x \oplus \beta ( f ( \tau (j) ) ) }
  = \Ket{j} \Ket{x + \beta ( f ( \tau (j) ) ) \mod 2^m}
  \label{eqn:bit-query}
\end{eqnarray}
For the rest of this paper we will refer to this query definition as the \emph{bit query}.

Consider the problem to compute the mean.
Let $F := \{ f : \{0, \ldots, N-1\} \rightarrow [0,1] \}$.
The solution operator $S_N : F \rightarrow [0,1] $ for this problem is
\begin{equation*}
  S_N (f) := \frac{1}{N} \sum_{j=0}^{N-1} f(j).
\end{equation*}
Brassard et al. \cite{bra-hoy-mos-tap-00}, Novak \cite{nov-01}, and Heinrich \cite{hei-01} showed that for this problem
\[ \compqeps (S_N, F) = \Theta ( \frac{1}{\epsilon} ) \]
with \emph{phase queries} as well as with \emph{bit queries}, where in both cases the upper bounds were shown by the amplitude estimation algorithm.

Does this equivalence hold for all problems?

\section{Bit queries can simulate phase queries}\label{sec:bit-simulates-phase}

Suppose we are able to construct a unitary transformation that realizes a phase query by the use of bit queries.
If we additionally show that the required number of bit queries is a small constant, no algorithm that uses phase queries could be (asymptotically) faster than an algorithm with bit queries.

\begin{defi}
  If it is possible to find unitary transformations $U_i$ not depending on $f$ with either
  \begin{eqnarray}
    \label{eqn:simulation}
    Q^{(1)}_f = U_{n_q} Q^{(2)}_f U_{n_q-1} \ldots U_{1} Q^{(2)}_f U_{0} \ \ \text{ (simulation) }
  \end{eqnarray}
  or
  \begin{eqnarray}
    \label{eqn:approximation}
    \left\| Q^{(1)}_f - U_{n_q} Q^{(2)}_f U_{n_q-1} \ldots U_{1} Q^{(2)}_f U_{0} \right\| \leq \delta \ \ \text{ (approximation) }
  \end{eqnarray}
  for all $f \in F$, we say that the query $Q_f^{(1)}$ is simulated or approximated up to $\delta$ by $Q_f^{(2)}$ using $n_q$ queries $Q_f^{(2)}$.
\end{defi}

What does this definition of approximation say about the probability to measure the correct state?
As it is shown e.g. in \cite{nie-chu-00}, box 4.1, suppose the probability that on input $\Ket{\psi}$, the state $\Ket{\varphi}$ is measured with probability $p_U$, if $U$ is applied, and with $p_V$, if $V$ is applied.
Then
\begin{equation}\label{eqn:approx-probability-error}
  | p_U - p_V | \leq 2 \| U - V \|.
\end{equation}

\begin{prop}
  Two applications of $Q_f^{\text{bit}}$ suffice to approximate $Q^{\text{phase}}_f$ by $Q_f^{\text{bit}}$.
  The precision of the approximation only depends on the encodings $\beta^{\text{phase}}$, $\beta^{\text{bit}}$, and the number of bits in the bit query $Q_f^{\text{bit}}$.
\end{prop}

Fix the encodings $\beta^{\text{bit}} : [0,1] \rightarrow \Set{0,\ldots, 2^m-1 }$, and $\beta^{\text{phase}} : [0,1] \rightarrow [0,1]$.
Define $\beta_-^{\text{bit}} : \{0, \ldots, 2^m-1\} \rightarrow [0,1]$ as a mapping that satisfies $\beta^{\text{bit}} \circ \beta_-^{\text{bit}} = \id$.

Let the starting state $\Ket{\psi}$ be
\[ \Ket{\psi} = \sum_j \alpha_j \Ket{j} \Ket{0} + \sum_j \beta_j \Ket{j} \Ket{1} \in H_n \otimes H_1. \]
We just state the algorithm for the basis states $\Ket{j} \Ket{0}$, and $\Ket{j} \Ket{1}$, the extension to arbitrary states $\Ket{\psi}$ follows from linearity.
\begin{itemize}
\item Starting state is $\Ket{j} \Ket{0}$ or $\Ket{j} \Ket{1}$ .
\item Append $\Ket{0} \Ket{0} \in H_n \otimes H_m$:
  \begin{eqnarray*}
    \Ket{j} \Ket{0} \Ket{0} \Ket{0} \ \ \text{ or } \ \ \Ket{j} \Ket{1} \Ket{0} \Ket{0}
  \end{eqnarray*}
\item Copy (by addition modulo $2^n$) the first register to the third register:
  \begin{eqnarray*}
    \Ket{j} \Ket{0} \Ket{j} \Ket{0} \ \ \text{ or } \ \ \Ket{j} \Ket{1} \Ket{j} \Ket{0}
  \end{eqnarray*}
\item Apply $Q_f^{\text{bit}}$ to the third and fourth register:
  \begin{align}\label{eqn:approx-query-applied}
    \Ket{j}\Ket{0}\Ket{j}\Ket{\beta^{\text{bit}}(f(\tau(j)))} \ \ \text{ or } \ \ 
    \Ket{j}\Ket{1}\Ket{j}\Ket{\beta^{\text{bit}}(f(\tau(j)))} 
  \end{align}
\item Define a mapping $U$ by
  \begin{equation}
    \label{eqn:approx-key-transform}
    \begin{split}
      U \Ket{j}\Ket{0}\Ket{j}\Ket{x}
      = &
      \sqrt{ 1 - \beta^{\text{phase}} ( \beta^{\text{bit}}_- (x) ) } \Ket{j} \Ket{0} \Ket{j} \Ket{x} \\
      & + \sqrt{ \beta^{\text{phase}} ( \beta^{\text{bit}}_- (x) ) } \Ket{j} \Ket{1} \Ket{j} \Ket{x} \\
      U \Ket{j}\Ket{1}\Ket{j}\Ket{x}
      = &
      - \sqrt{ \beta^{\text{phase}} ( \beta^{\text{bit}}_- (x) ) } \Ket{j} \Ket{0} \Ket{j} \Ket{x} \\
      & + \sqrt{ 1 - \beta^{\text{phase}} ( \beta^{\text{bit}}_- (x) ) } \Ket{j} \Ket{1} \Ket{j} \Ket{x}
    \end{split}
  \end{equation}
  for every $j \in \Set{0,\ldots,2^n-1}$ and $x \in \Set{0,\ldots,2^m-1}$.
  Note that $U$ does not depend on $f$.
  Apply $U$ to the state (\ref{eqn:approx-query-applied}), resulting in:
  \begin{multline}\label{eqn:approx-final-0}
    \sqrt{ 1 - \beta^{\text{phase}} ( \beta^{\text{bit}}_- ( \beta^{\text{bit}}(f(\tau(j))) ) ) } \Ket{j} \Ket{0} \Ket{j} \Ket{\beta^{\text{bit}}(f(\tau(j)))} \\
    + \sqrt{ \beta^{\text{phase}} ( \beta^{\text{bit}}_- ( \beta^{\text{bit}}(f(\tau(j))) ) ) } \Ket{j} \Ket{1} \Ket{j} \Ket{\beta^{\text{bit}}(f(\tau(j)))}
  \end{multline}
  or
  \begin{multline}\label{eqn:approx-final-1}
    - \sqrt{ \beta^{\text{phase}} ( \beta^{\text{bit}}_- ( \beta^{\text{bit}}(f(\tau(j))) ) ) } \Ket{j} \Ket{0} \Ket{j} \Ket{\beta^{\text{bit}}(f(\tau(j)))} \\
    + \sqrt{ 1 - \beta^{\text{phase}} ( \beta^{\text{bit}}_- ( \beta^{\text{bit}}(f(\tau(j))) ) ) } \Ket{j} \Ket{1} \Ket{j} \Ket{\beta^{\text{bit}}(f(\tau(j)))}.
  \end{multline}
\item Uncompute $Q_f^{\text{bit}}$ by mapping
  \begin{eqnarray*}
    \Ket{j} \Ket{b} \Ket{k} \Ket{x} \mapsto \Ket{j} \Ket{b} \Ket{k} \Ket{(- x) \mod 2^m}
  \end{eqnarray*}
  and applying $Q_f^{\text{bit}}$.
  Now map
  \begin{eqnarray*}
    \Ket{j} \Ket{b} \Ket{k} \Ket{x} \mapsto \Ket{j} \Ket{b} \Ket{(-k) \mod 2^n} \Ket{x}
  \end{eqnarray*}
  and again apply the copy operation.
  We can discard the additional quantum registers now, since they are in the state $\Ket{0} \Ket{0}$ again.
\end{itemize}

A close look at the states in (\ref{eqn:approx-final-0}), and (\ref{eqn:approx-final-1}) reveals that we have successfully simulated the mapping
\[ Q^{\text{phase}}_{\beta^{\text{bit}}_- \circ \beta^{\text{bit}} \circ f}. \]
Since in general $\beta^{\text{bit}}_- \circ \beta^{\text{bit}} \neq \id$, our approximation introduced an error in the simulation.
This error is bounded by
\begin{align}
  \label{eqn:approx-error}
  \sup_{x \in D} | \beta^{\text{bit}}_- ( \beta^{\text{bit}} ( f ( x ) ) ) - f(x) |
  & \leq \sup_{y \in [0,1]} | \beta^{\text{bit}}_- ( \beta^{\text{bit}} ( y ) ) - y | \nonumber\\
  & \leq \sup \Set{ | x - y | \, | \, \beta^{\text{bit}} (x) = \beta^{\text{bit}} (y)}.
\end{align}
For the most obvious choice for $\beta^{\text{bit}}$
\begin{equation}
  \label{eqn:floor-beta-def}
  \begin{split}
    \beta^{\text{bit}} : [0,1) & \rightarrow \Set{0, \ldots, 2^m-1} \\
    x & \mapsto \lfloor x \cdot 2^m \rfloor
  \end{split}
\end{equation}
a good $\beta^{\text{bit}}_-$ is
\begin{equation}
  \label{eqn:floor-beta--def}
  \begin{split}
    \beta^{\text{bit}}_- :  \Set{0, \ldots, 2^m-1} & \rightarrow [0,1] \\
    x & \mapsto  x \cdot 2^{-m} + 2^{-m-1}
  \end{split}
\end{equation}
for which the error from (\ref{eqn:approx-error}) is less than $2^{-m-1}$.

As an example consider $\beta^{\text{phase}} = \id$.
After some calculation we get
\begin{equation*}
  \| Q_f^{\text{phase}} - Q_{\beta^{\text{bit}}_- \circ \beta^{\text{bit}} \circ f}^{\text{phase}} \| \leq 2^{-m/2}.
\end{equation*}

\section{Phase queries cannot efficiently approximate bit queries}

Let us consider the (simple) evaluation problem
\begin{equation}
  \label{eqn:evaluation-problem}
  \begin{split}
    S : F & \rightarrow [0,1] \\
    f & \mapsto S f := f(0)
  \end{split}
\end{equation}
with $F = \{ f : \{ 0 \} \rightarrow [0,1] \}$.
It is obvious that this problem can be solved on a ``classical'' computer by just one evaluation of $f$, i.e. $\compeps (S,F) = 1$.
The precision $\epsilon$ only depends on the accuracy of the query - the function call returning $f(0)$.

On a quantum computer that uses bit queries to evaluate $f$ also
\begin{equation}
  \label{eqn:evaluation-bound-bit}
  \compqbeps (S,F) = 1,
\end{equation}
at least, if we choose $\beta^{\text{bit}}$ as in (\ref{eqn:floor-beta-def}) and if the number of qubits $m$ is chosen so that $2^m > \frac{1}{\epsilon}$.

If we are only allowed to use phase queries however, we can prove a lower bound theorem.
\begin{theo}\label{theo:evaluation-phase-bound}
  Consider the evaluation problem (\ref{eqn:evaluation-problem}).
  For any quantum algorithm that is only allowed to access $f$ by phase queries,
  \begin{equation}
    \label{eqn:evaluation-bound-phase}
    \compqpeps (S,F) = \Omega ( \frac{1}{\epsilon} ),
  \end{equation}
  if the encoding $\beta^{\text{phase}}$ is chosen as $\beta^{\text{phase}} = \id$.
\end{theo}

To prove this theorem we will need lemma \ref{lemm:quant-algo-trig-polys} and lemma \ref{lemm:trig-poly-bound}, thus we will postpone the proof.
But theorem \ref{theo:evaluation-phase-bound} allows us to easily prove the following corollary:
\begin{corr}
  The approximation of an $m$-bit query $Q_f^{\text{bit}}$ to $\delta \leq \frac{1}{8}$ by a phase query $Q_f^{\text{phase}}$ requires $\Omega (2^m)$ phase queries $Q_f^{\text{phase}}$ when $\beta^{\text{phase}} = \id$.
\end{corr}
\begin{proof}
  Suppose we could approximate an $m$-bit query by $n_q = n_q (m)$ phase queries and for all $c > 0$, $m_0 \in \mathbb{N}$ there exists an $m \geq m_0$ with
  \begin{equation*}
    n_q (m) < c 2^m.
  \end{equation*}
  Fix such an $m$.
  Approximation means
  \begin{equation*}
    \left\| Q^{\text{bit}}_f - U_{n_q} Q^{\text{phase}}_f U_{n_q-1} \ldots U_{1} Q^{\text{phase}}_f U_{0} \right\| \leq \delta.
  \end{equation*}
  We know that just one $m$-bit query is sufficient to solve the evaluation problem to accuracy $\epsilon := 2^{-m+1}$ and certainty, so we could use the following algorithm:
  \begin{equation*}
    \widetilde{U}_1 U_{n_q} Q^{\text{phase}}_f U_{n_q-1} \ldots U_{1} Q^{\text{phase}}_f U_{0} \widetilde{U}_0 \Ket{\psi},
  \end{equation*}
  where
  \begin{equation*}
    \widetilde{U}_1 Q^{\text{bit}}_f \widetilde{U}_0 \Ket{\psi}    
  \end{equation*}
  is the algorithm that uses just one bit query.

  The probability that the output of this algorithm with an approximated query is incorrect would be less or equal than $2 \delta$, see (\ref{eqn:approx-probability-error}).
  Thus we could solve the evaluation problem with
  \begin{equation*}
    \compqeps (S,F) \leq n_q < c 2^{m} = 2 c \frac{1}{\epsilon},
  \end{equation*}
  which is a contradiction to theorem \ref{theo:evaluation-phase-bound}.
\end{proof}

With the result from section \ref{sec:bit-simulates-phase} that any phase query can be approximated by 2 bit queries we can conclude that bit queries are more powerful than phase queries.

To prove theorem \ref{theo:evaluation-phase-bound} let us introduce multivariate trigonometric polynomials (for an introduction see e.g. \cite{nik-75}).

\begin{defi}\label{defi:multivariate-trig-polys}
  A mapping $T : \mathbb{R}^n \rightarrow \mathbb{C}$ is called an \textbf{$n$-variate trigonometric polynomial}, if it is of the form
  \[ T ( \theta_0, \ldots, \theta_{n-1} ) = \sum_{j \in J} c_j e^{i (n_{j,0} \theta_0 + \ldots + n_{j,n-1} \theta_{n-1}) } \]
  with $c_j \in \mathbb{C}$, $n_{j,k} \in \mathbb{Z}$, and $|J| < \infty$.
  We additionally define
  \[ \deg T( \theta_0, \ldots, \theta_{n-1} ) := \max_{j \in J} \left( | n_{j,0} | + \ldots + | n_{j,n-1} | \right) \]
  as the \textbf{degree} of $T( \theta_0, \ldots, \theta_{n-1} )$.
\end{defi}

This definition allows us to define for every quantum algorithm, which depends on a phase query $Q_f^{\text{phase}}$, a corresponding multivariate polynomial.
The idea that allows this is that every phase query as in (\ref{eqn:phase-query}) can also be written as
\begin{eqnarray}
  \label{eqn:phase-query-as-rotation}
  Q^{\text{phase}}_f \Ket{j} \Ket{0}
  & = & \cos \theta_j \Ket{j} \Ket{0} + \sin \theta_j \Ket{j} \Ket{1} \nonumber\\
  Q^{\text{phase}}_f \Ket{j} \Ket{1}
  & = & - \sin \theta_j\Ket{j} \Ket{0} + \cos \theta_j \Ket{j} \Ket{1},
\end{eqnarray}
if $\theta_j$ is chosen so that
\begin{equation}
  \label{eqn:theta-j-def}
  \theta_j = \arcsin \sqrt{\beta^{\text{phase}}(f(\tau(j)))}.
\end{equation}
Now we can prove the following lemma, which closely follows the proof idea in Beals et al., \cite{bea-buh-cle-mos-wol-98}.

\begin{lemm}\label{lemm:quant-algo-trig-polys} 
  For every quantum algorithm $A$ making $n_q$ queries $Q_f^{\text{phase}}$ there are trigonometric polynomials $T_k( \theta_0, \ldots, \theta_{n-1} )$, $\deg T_k( \theta_0, \ldots, \theta_{n-1} ) \leq n_q$, so that
  \begin{equation*}
    \label{eqn:quant-algo-trig-poly}
    \begin{split}
      A \Ket{\psi} 
      & = U_{n_q} Q_f^{\text{phase}} U_{n_q-1} \ldots U_1 Q_f^{\text{phase}} U_0 \Ket{\psi} \\
      & = \sum_k T_k(\theta_0, \ldots, \theta_{n-1}) \Ket{k}
    \end{split}
  \end{equation*}
  holds with $\theta_j$ defined as in (\ref{eqn:theta-j-def}).
\end{lemm}
\begin{proof}
  The proof is by induction over the number of queries ${n_q}$.
  \medskip
  \begin{flushleft}
    \underline{$n_q=0:$} we have not made any queries $Q_f^{\text{phase}}$ therefore the state can be written as
    \begin{equation*}
      U_0 \Ket{\psi} = \sum_k \alpha_k \Ket{k}
    \end{equation*}
    with constants $\alpha_k$.
  \end{flushleft}
  \medskip
  \begin{flushleft}
    \underline{$n_q \mapsto n_q + 1 :$} suppose
    \begin{equation*}
      U_{n_q} Q_f^{\text{phase}} U_{n_q-1} \ldots U_1 Q_f^{\text{phase}} U_0 \Ket{\psi} 
      = \sum_k T_k(\theta_0, \ldots, \theta_{n-1}) \Ket{k}.
    \end{equation*}
    with $ \deg T_{k} (\theta_0, \ldots, \theta_{n-1}) \leq {n_q}$.
    We can split this state into
    \begin{equation}
      \label{eqn:pre-query}
      \begin{split}
        \sum_k T_k(\theta_0, \ldots, \theta_{n-1}) \Ket{k}
        = & \sum_{i,j,l} T_{ij0l} (\theta_0, \ldots, \theta_{n-1}) \Ket{i} \Ket{j} \Ket{0} \Ket{l} \\
        + & \sum_{i,j,l} T_{ij1l} (\theta_0, \ldots, \theta_{n-1}) \Ket{i} \Ket{j} \Ket{1} \Ket{l}.
      \end{split}
    \end{equation}
    Let $\Theta = (\theta_0, \ldots, \theta_{n-1})$.
    After applying $Q_f^{\text{phase}}$ to state (\ref{eqn:pre-query}), the result can be written as
    \begin{eqnarray*}
      &   & \sum_{i,j,l} \Ket{i} \left( \cos \theta_j T_{ij0l} (\Theta) \Ket{j}\Ket{0} + \sin \theta_j T_{ij0l} (\Theta) \Ket{j}\Ket{1} \right) \Ket{l} \\
      & + & \sum_{i,j,l} \Ket{i} \left( - \sin \theta_j T_{ij1l} (\Theta) \Ket{j}\Ket{0} + \cos \theta_j T_{ij1l} (\Theta) \Ket{j}\Ket{1} \right) \Ket{l} \\
      & = & \sum_{i,j,l} \left( \cos \theta_j T_{ij0l} (\Theta) - \sin \theta_j T_{ij1l} (\Theta) \right) \Ket{i}\Ket{j}\Ket{0}\Ket{l} \\
      & + & \sum_{i,j,l} \left( \sin \theta_j T_{ij0l} (\Theta) + \cos \theta_j T_{ij1l} (\Theta) \right) \Ket{i}\Ket{j}\Ket{1}\Ket{l}.
    \end{eqnarray*}
  \end{flushleft}
  
  The terms
  \[ \cos \theta_j T_{ij0l} ( \theta_0, \ldots, \theta_{n-1} ) - \sin \theta_j T_{ij1l} ( \theta_0, \ldots, \theta_{n-1} ) \]
  and
  \[ \sin \theta_j T_{ij0l} ( \theta_0, \ldots, \theta_{n-1} ) + \cos \theta_j T_{ij1l} ( \theta_0, \ldots, \theta_{n-1} ) \]
  again are multivariate trigonometric polynomials and of degree at most $n_q + 1$.

  Finally the mapping $U_{{n_q}+1}$ is independent of $f$, thus not dependent on $\Theta$, and just maps to a new linear combination of trigonometric polynomials, which again has degree at most $n_q+1$.    
\end{proof}

For (univariate) trigonometric polynomials we cite (see e.g. \cite{che-66}, \cite{riv-74})

\begin{theo}[Bernstein's Inequality]\label{theo:bernstein-inequality}
  Let $t(\theta)$ be a trigonometric polynomial.
  Then
  \[ \max_{-\pi \leq \theta \leq \pi} |t'(\theta)| \leq \deg t(\theta) \max_{-\pi \leq \theta \leq \pi} | t(\theta) | . \]
\end{theo}

The trigonometric polynomials $T_k (\theta_0, \ldots, \theta_{n-1})$ we found in lemma \ref{lemm:quant-algo-trig-polys} were related to $f$ via equation (\ref{eqn:theta-j-def}):
\begin{equation*}
  \theta_j = \arcsin \sqrt{\beta^{\text{phase}}(f(\tau(j)))}.
\end{equation*}
What is the bound of Bernstein's inequality in dependence on $\arcsin \sqrt{\cdot}$?

\begin{lemm}\label{lemm:trig-poly-bound}
  If there exist an $x \in [0,1]$, and a $\Delta$ with $x + \Delta \in [0,1]$ so that for the trigonometric polynomial $t(\theta)$
  \[ t( \arcsin \sqrt{x} ) \geq \frac{3}{4} \ \text{ and } \ t( \arcsin \sqrt{x + \Delta} ) \leq \frac{1}{4} \]
  and $t(\theta) \in [0,1]$ for all $\theta \in \mathbb{R}$, then
  \begin{equation}
    \label{eqn:degree-bound-from-lemma}
    \deg t(\theta) \geq c \left( \sqrt{ \frac{1}{ |\Delta| } } + \frac{ \sqrt{m (1-m)} }{ |\Delta| } \right),
  \end{equation}
  with $m \in \{ x, x + \Delta \}$ so that $| m - \frac{1}{2} |$ is maximal and a constant $c$.  
\end{lemm}
\begin{proof}
  Applying the mean value theorem yields a $\xi \in [x, x + \Delta]$ or $\xi \in [x+\Delta,x]$ with
  \begin{equation*}
    | t(\arcsin \sqrt{x} ) - t(\arcsin \sqrt{x+\Delta}) |
    = \left| \arcsin \sqrt{x} - \arcsin \sqrt{x+\Delta} \right| | t' (\xi) | 
  \end{equation*}
  and thus
  \begin{equation*}
    | t' (\xi) | \geq \frac{1}{2} \left| \arcsin \sqrt{x} - \arcsin \sqrt{x+\Delta} \right|^{-1}.
  \end{equation*}
  Now bound the degree of $t(\theta)$ by Bernstein's inequality (lemma \ref{theo:bernstein-inequality}) and use that $t(\theta) \in [0,1]$:
  \begin{equation}
    \label{eqn:deg-arcsin-bound}
    \frac{1}{2} \left| \arcsin \sqrt{x+\Delta} - \arcsin \sqrt{x} \right|^{-1}
    \leq \deg t (\theta) \max_{-\pi \leq \theta \leq \pi} | t(\theta) |
    \leq \deg t (\theta).
  \end{equation}

  For the first bound we use that for all $\varphi, \psi \in [0,\pi/2]$
  \begin{equation}
    \label{eqn:sin-square-inequality}
    \frac{2}{\pi} | \varphi - \psi |
    \leq \sqrt{ 2 | \sin^2 \varphi - \sin^2 \psi | },
  \end{equation}
  and therefore from (\ref{eqn:deg-arcsin-bound}) and (\ref{eqn:sin-square-inequality})
  \begin{equation*}
    \deg t(\theta) 
    \geq c \sqrt{ \frac{1}{ \left| \sin^2 \arcsin \sqrt{x+\Delta} - \sin^2 \arcsin \sqrt{x} \right| } }
    = c \sqrt{\frac{1}{|\Delta|}}.
  \end{equation*}
  
  The second bound is proven by another application of the mean-value theorem.
  There exists a $\xi \in [x, x + \Delta]$ or $\xi \in [x+\Delta,x]$ with
  \begin{equation*}
    \deg t(\theta)
    \geq \frac{1}{2} \left| ( \arcsin \sqrt{\xi} )' ( x + \Delta - x ) \right|^{-1}
  \end{equation*}
  Note that for $\xi \in [x, x + \Delta]$ or $\xi \in [x+\Delta,x]$
  \begin{equation*}
    (\arcsin \sqrt{\xi})' = \frac{1}{2 \sqrt{\xi(1-\xi)}}
  \end{equation*}
  is maximized by $\xi = m$, with $m \in \Set{x, x+\Delta}$ so that $|m-\frac{1}{2}|$ is maximal.
  Thus
  \begin{equation*}
    \deg t(\theta)
    \geq c \frac{ \sqrt{m (1-m)} }{ |\Delta| }.
  \end{equation*}
  
\end{proof}

Note that the bound (\ref{eqn:degree-bound-from-lemma}) closely resembles the bound of Nayak and Wu, \cite{nay-wu-99}.
With the help of lemma \ref{lemm:trig-poly-bound} we are able to prove theorem \ref{theo:evaluation-phase-bound}.
\begin{proof}[Proof of theorem \ref{theo:evaluation-phase-bound}]
  Let $n_q$ be the number of queries of an algorithm for the evaluation problem.
  By lemma \ref{lemm:quant-algo-trig-polys} trigonometric polynomials $T_k (\theta)$ exist with
  \begin{equation}\label{eqn:final-quantum-state}
    U_{n_q} Q^{\text{phase}}_f \ldots U_{0} \Ket{\psi}
    = \sum_{k} T_{k} ( \theta ) \Ket{k},
  \end{equation}
  for $\theta = \arcsin \sqrt{\beta^{\text{phase}}(f(\tau(j)))}$, see (\ref{eqn:theta-j-def}), and $\deg T_k (\theta) \leq n_q$.

  The state (\ref{eqn:final-quantum-state}) is measured and $\varphi$ maps the result of the measurement to an element of our solution set $[0,1]$.
  Let $B$ be the set of all states so that
  \begin{equation}\label{eqn:states-eps-close-to-1/2}
    B := \Set{ k \, | \, | \varphi(k) - 1/2 | < \epsilon}.
  \end{equation}
  The probability to measure a state $\Ket{k}$ is given by $| T_k (\theta) |^2$.
  Let
  \begin{equation*}
    T_k (\theta) = \sum_l \alpha_{k,l} e^{i n_{k,l} \theta}.
  \end{equation*}
  Then
  \begin{equation*}
    \sum_{k \in B} \left| T_k (\theta) \right|^2
    = \sum_{k \in B} \left( \sum_l \alpha_{k,l} e^{i n_{k,l} \theta} \right) \left( \sum_l \overline{\alpha_{k,l}} e^{- i n_{k,l} \theta} \right) 
    =: T(\theta),
  \end{equation*}
  which is a trigonometric polynomial again, $\deg T (\theta) \leq 2 n_q$, and $T(\theta) \in [0,1]$.
  
  Choose the input $f_1, f_2 \in F$ with $f_1 (0) = \frac{1}{2}$, $f_2 (0) = \frac{1}{2} - 2 \epsilon$, $0 < \epsilon < \frac{1}{4}$.
  If the algorithm is correct,
  \begin{equation*}
    T (\arcsin \sqrt{ \beta^{\text{phase}}( f_1(0) ) }) \geq \frac{3}{4} \ \ \text{ and } \ \
    T (\arcsin \sqrt{ \beta^{\text{phase}}( f_2(0) ) }) \leq \frac{1}{4}
  \end{equation*}
  (all query inputs are mapped to $0$ by $\tau$).

  Our lemma \ref{lemm:trig-poly-bound} tells us a bound for the degree of $T(\theta)$:
  \begin{equation*}
    2 n_q \geq \deg T (\theta) \geq c \left( \sqrt{ \frac{1}{ |\Delta| } } + \frac{ \sqrt{m (1-m)} }{ |\Delta| } \right),
  \end{equation*}
  where
  \begin{equation}
    \label{eqn:beta-gives-Delta}
    \Delta = \beta^{\text{phase}}( f_1(0) ) - \beta^{\text{phase}}( f_2(0) ).
  \end{equation}
  and $m \in \Set{ \beta^{\text{phase}}( f_1(0) ), \beta^{\text{phase}}( f_2(0) )}$ so that $| m - \frac{1}{2} |$ is maximal.

  We chose $\beta^{\text{phase}} = \id$, so
  \begin{equation*}
    2 n_q
    \geq c \left( \sqrt{ \frac{1}{ 2 \epsilon } } + \frac{ \sqrt{ (\frac{1}{2} - 2 \epsilon )( 1 - \frac{1}{2} + 2 \epsilon) } }{ 2 \epsilon } \right)
    = c \left( \sqrt{ \frac{1}{ 2 \epsilon } } + \frac{ \sqrt{ \frac{1}{4}  - 4 \epsilon^2} }{ 2 \epsilon } \right).
  \end{equation*}
  
  There exists a constant $c'$ so that for $\epsilon$ small enough
  \begin{equation*}
    n_q \geq c' \frac{1}{\epsilon}
  \end{equation*}
  holds, therefore
  \begin{equation*}
    \compqpeps (S,F) = \Omega( \frac{1}{\epsilon} ).
  \end{equation*}
\end{proof}

\section{Another proof for the hardness of evaluation with phase queries}

An anonymous referee suggested a shorter proof for theorem \ref{theo:evaluation-phase-bound}.
The proof is based on the observation that for two inputs $f_1$ and $f_2$ which are chosen as in the proof of theorem \ref{theo:evaluation-phase-bound}, $f_1, f_2 \in F = \{ f : \{0\} \rightarrow [0,1] \}$,
\begin{equation*}
  f_1 (0) = \frac{1}{2} \ \text{ and } \ f_2 (0) = \frac{1}{2} - 2 \epsilon ,
\end{equation*}
the difference of $Q_{f_1}^{\text{phase}}$ and $Q_{f_2}^{\text{phase}}$ is bounded in the operator norm by
\begin{equation*}
  \begin{split}
    & \; \| Q_{f_1}^{\text{phase}} - Q_{f_2}^{\text{phase}} \|
    = \sup_{\Ket{x}, \| \Ket{x} \| = 1} \| (Q_{f_1}^{\text{phase}} - Q_{f_2}^{\text{phase}}) \Ket{x} \| \\
    = & \; \sup \| (Q_{f_1}^{\text{phase}} - Q_{f_2}^{\text{phase}}) ( a_{00} \Ket{0} \Ket{0} + a_{01} \Ket{0} \Ket{1} \|
  \end{split}
\end{equation*}
with $a_{00}, a_{01} \in \mathbb{C}$ and $|a_{00}|^2+|a_{01}|^2 = 1$, since $Q_{f_1}^{\text{phase}} - Q_{f_2}^{\text{phase}} = 0$ for all other inputs.

Each of the terms can be bounded in the following way (here for $a_{00} \Ket{0}\Ket{0}$):
\begin{equation*}
  \begin{split}
    & \; \| (Q_{f_1}^{\text{phase}} - Q_{f_2}^{\text{phase}}) a_{00} \Ket{0} \Ket{0} \| \\
    = & \; |a_{00}| \Norm{ \sqrt{ 1/2 } \Ket{0} \Ket{0} + \sqrt{ 1/2 } \Ket{0} \Ket{1}
    - \sqrt{ 1/2 + 2 \epsilon } \Ket{0} \Ket{0} - \sqrt{ 1/2 - 2 \epsilon } \Ket{0} \Ket{1} } \\
    = & \; \frac{|a_{00}| }{\sqrt{2}} \Norm{ (1-\sqrt{1+4 \epsilon}) \Ket{0}
      + (1-\sqrt{1-4\epsilon}) \Ket{1} } \\
    = & \; \frac{|a_{00}| }{\sqrt{2}} \sqrt{ 1 - 2 \sqrt{1+4\epsilon} + 1+4\epsilon
      + 1 - 2 \sqrt{1-4\epsilon} + 1-4\epsilon } \\
    = & \; |a_{00}| \sqrt{ 2 - \sqrt{1+4\epsilon} - \sqrt{1-4\epsilon} }.
  \end{split}
\end{equation*}

If we develop $\sqrt{ 1 + 4 x }$ into a Taylor series we get
\begin{equation*}
\sqrt{ 1 + 4 x } = 1 + 2 x - 2 x^2 \pm \mathcal{O} (|x|^3)
\end{equation*}
and thus
\begin{equation*}
  \begin{split}
    & \; \| (Q_{f_1}^{\text{phase}} - Q_{f_2}^{\text{phase}}) a_{00} \Ket{0} \Ket{0} \| \\
    \leq & \; |a_{00}| \sqrt{ 2 - 2 + 4 \epsilon^2 \mp \mathcal{O} ( |\epsilon|^3 ) } \\
    = & \; |a_{00}| \, \epsilon \, \sqrt{4 \mp \mathcal{O} (|\epsilon|) } \\
    \in & \; |a_{00}| \mathcal{O} (\epsilon) .
  \end{split}
\end{equation*}
The bounds for the term $a_{01} \Ket{0}\Ket{1}$ are analogous.

Therefore we can conclude that
\begin{equation*}
  \| Q_{f_1}^{\text{phase}} - Q_{f_2}^{\text{phase}} \| \in \mathcal{O} (\epsilon).
\end{equation*}

Now we turn to our quantum algorithm, which is of the form described in equation (\ref{eqn:quantum-algorithm}):
\begin{equation*}
  U_n Q_f^{\text{phase}} U_{n-1} \ldots U_1 Q_f^{\text{phase}} U_0 \Ket{\psi}.
\end{equation*}
We would like to bound the difference between the algorithm on input $f_1$ and on $f_2$:
\begin{equation}\label{eqn:norm-difference-f1-f2}
  \Norm{ U_n Q_{f_1}^{\text{phase}} U_{n-1} \ldots U_1 Q_{f_1}^{\text{phase}} U_0 \Ket{\psi}
    - U_n Q_{f_2}^{\text{phase}} U_{n-1} \ldots U_1 Q_{f_2}^{\text{phase}} U_0 \Ket{\psi} }.
\end{equation}
We will use the following inequality for unitary matrices $A$, $B$, $C$, and $D$:
\begin{equation}\label{eqn:mat-prod-diff-bound}
  \begin{split}
    & \; \Norm{ A B - C D}
    = \Norm{ A B - C B + C B - C D } \\
    \leq & \; \Norm{ A B - C B } + \Norm{ C B - C D}
    \leq \Norm{A-C} \Norm{B} + \Norm{C} \Norm{B-D} \\
    = & \; \Norm{A-C} + \Norm{B-D}
  \end{split}
\end{equation}
(see e.g. Nielsen and Chuang \cite{nie-chu-00}).
Hence equation (\ref{eqn:norm-difference-f1-f2}) is bounded by
\begin{equation*}
  n \Norm{ Q_{f_1}^{\text{phase}} - Q_{f_2}^{\text{phase}} }.
\end{equation*}

Recall the definition of $B$, the set of all states representing $\frac{1}{2}$ from equation (\ref{eqn:states-eps-close-to-1/2}).
Our algorithm for the evaluation problem must ensure that for $f_1$ the probability to measure a state from $B$ is greater than $\frac{3}{4}$ and for $f_2$ less than $\frac{1}{4}$.
Let $p_{f_1}$ be the probability to measure a state from $B$ on input $f_1$ and $p_{f_2}$ be the probability on input $f_2$.
We want to have
\begin{equation}\label{eqn:prob-difference-reqirement}
  | p_{f_1} - p_{f_2} | \geq \frac{1}{2}.
\end{equation}

We use the fact from equation (\ref{eqn:approx-probability-error}) that for $A_{f_1} := U_n Q_{f_1}^{\text{phase}} \ldots U_0$ and $A_{f_2} := U_n Q_{f_2}^{\text{phase}} \ldots U_0$ and a measurement projection $P$
\begin{equation*}
  \begin{split}
    & \; | p_{f_1} - p_{f_2} |
    = \left| \Bra{\psi} A_{f_1}^{\dag} P A_{f_1} \Ket{\psi}
      - \Bra{\psi} A_{f_2}^{\dag} P A_{f_2} \Ket{\psi} \right| \\
    = & \; \left| \Bra{\psi} \left( 
        A_{f_1}^{\dag} P (A_{f_1}-A_{f_2})
        + A_{f_1}^{\dag} P A_{f_2}
        + (A_{f_1}^{\dag} - A_{f_2}^{\dag}) P A_{f_2}
        - A_{f_1}^{\dag} P A_{f_2} \right) \Ket{\psi} \right| \\
    \leq & \; \left| \Bra{\psi} A_{f_1}^{\dag} P (A_{f_1}-A_{f_2}) \Ket{\psi} \right|
    + \left| \Bra{\psi} (A_{f_1}^{\dag} - A_{f_2}^{\dag}) P A_{f_2} \Ket{\psi} \right| \\
    \leq & \; \Norm{P A_{f_1} \Ket{\psi}} \Norm{(A_{f_1}-A_{f_2}) \Ket{\psi}}
    + \Norm{(A_{f_1} - A_{f_2}) \Ket{\psi}} \Norm{P A_{f_2} \Ket{\psi}} \\
    \leq & \; 2 \Norm{ A_{f_1} - A_{f_2} } ,
  \end{split}
\end{equation*}
where we used the Cauchy-Schwarz inequality.
This bound can be found, e.g., in Nielsen and Chuang \cite{nie-chu-00}.

We choose $P$ as a projection on the subspace $\spanSpace B$ and get
\begin{equation}\label{eqn:alternate-main-inequality}
  \begin{split}
    | p_{f_1} - p_{f_2} |
    \leq & \; 2 \Norm{ A_{f_1} - A_{f_2} } \\
    \leq & \; 2 n \Norm{Q_{f_1}^{\text{phase}} - Q_{f_2}^{\text{phase}}} \\
    \in & \; 2 n \mathcal{O} (\epsilon)
  \end{split}
\end{equation}

Combining equations (\ref{eqn:prob-difference-reqirement}) and (\ref{eqn:alternate-main-inequality}) now yields
\begin{equation*}
  n \in \Omega \left( \frac{1}{\epsilon} \right).
\end{equation*}

\section{Acknowledgments}

This work was done as a master thesis under the supervision of Stefan Heinrich.
I am very grateful for his guidance and the many helpful discussions we had on quantum computing and complexity theory.
His valuable suggestions were greatly appreciated.

\bibliographystyle{plain}
\bibliography{qc}

\end{document}